\documentclass[twocolumn,pre]{revtex4}

\usepackage{graphicx}
\usepackage{color}
\usepackage{rotating}
\usepackage{amsmath}
\usepackage{amsfonts}
\usepackage{amssymb}
\usepackage{enumerate}
\usepackage{longtable}
\usepackage{dcolumn}
\usepackage{bm}
\usepackage{epsfig}
\usepackage{hyperref}
\begin{document}

\bibliographystyle{apsrev4-1}

\title{Emergence of Bursts and Communities in Evolving Weighted Networks}

\author{Hang-Hyun Jo}
\affiliation{BECS, Aalto University School of Science, P.O. Box 12200, FI-00076}
\author{Raj Kumar Pan}
\affiliation{BECS, Aalto University School of Science, P.O. Box 12200, FI-00076}
\author{Kimmo Kaski}
\affiliation{BECS, Aalto University School of Science, P.O. Box 12200, FI-00076}

\begin{abstract}
Understanding the patterns of human dynamics and social interaction, and the way they lead to the formation of an organized and functional society are important issues especially for techno-social development. Addressing these issues of social networks has recently become possible through large scale data analysis of e.g. mobile phone call records, which has revealed the existence of modular or community structure with many links between nodes of the same community and relatively few links between nodes of different communities. The weights of links, e.g. the number of calls between two users, and the network topology are found correlated such that intra-community links are stronger compared to the weak inter-community links. This feature is known as Granovetter's ``The strength of weak ties" hypothesis. In addition to this inhomogeneous community structure, the temporal patterns of human dynamics turn out to be inhomogeneous or bursty, characterized by the heavy tailed distribution of time interval between two consecutive events, i.e. inter-event time. In this paper, we study how the community structure and the bursty dynamics emerge together in a simple evolving weighted network model. The principal mechanisms behind these patterns are social interaction by cyclic closure, i.e. links to friends of friends and the focal closure, i.e. links to individuals sharing similar attributes or interests, and human dynamics by task handling process. These three mechanisms have been implemented as a network model with local attachment, global attachment, and priority-based queuing processes. By comprehensive numerical simulations we show that the interplay of these mechanisms leads to the emergence of heavy tailed inter-event time distribution and the evolution of Granovetter-type community structure. Moreover, the numerical results are found to be in qualitative agreement with empirical analysis results from mobile phone call dataset. 
\end{abstract}

\maketitle

\section{Introduction}

Human dynamics and social interaction patterns have been a subject of intensive study in many different fields ranging from sociology and economics to computer science and statistical physics constituting what is nowadays called network science~\cite{Wasserman1994,Goyal2007,Newman2010,Barabasi2010}. Partially due to the fact that huge amounts of various kinds of digital data on human dynamics have become available, explorative and quantitative analysis of these kinds of data has enabled us to have unprecedented insight into the structure and dynamics of behavioral, social, and even societal patterns. Examples of such data consist of email correspondence~\cite{Eckmann2004,Barabasi2005}, mobile phone call (MPC) and Short Message (SM) communication~\cite{Onnela2007a,Gonzalez2008,Wu2010}, online social network services~\cite{Kwak2010,Onnela2010}, and scientific collaboration~\cite{Newman2001}. 

The interaction structure among individuals in such large scale social data has been investigated by applying the concepts and methods of complex networks where individuals and their relationships represent nodes and links, respectively~\cite{Albert2002,Newman2003,Boccaletti2006}. In many real networks, the link is characterized by a weight corresponding to the strength or closeness of social relationship~\cite{Barrat2004,Boccaletti2006}, which in the case of MPC can be described by the aggregate number of calls between two individuals~\cite{Onnela2007a,Onnela2007b}. It has turned out that social networks are inhomogeneous and they can be characterized by modular or community structure~\cite{Fortunato2010}: The whole network is composed of separate communities connected by bridges, i.e. there are more and stronger links within communities than between communities, in accordance with Granovetter's ``The strength of weak ties" hypothesis~\cite{Granovetter1973}, corroborated later in~\cite{Onnela2007a,Onnela2007b}. This weight-topology coupling was successfully reproduced in the model of weighted networks driven by the cyclic and the focal closure processes~\cite{Kumpula2007}. Here the cyclic closure process refers to the link formation with one's next nearest neighbors, i.e. the link formation with friends of friends. The focal closure refers to the attribute-related link formation independently of the local connectivity~\cite{Kossinets2006}. It has been shown that these simple processes can lead to the emergence of complex weight-topology coupling, where the inhomogeneity of weights is a crucial factor for the emergence of communities.

In addition to the inhomogeneous community structure of social networks, the temporal patterns of human dynamics are inhomogeneous or bursty~\cite{Barabasi2005,Vazquez2006,Karsai2011}. The bursts of rapidly occurring events of activity are separated by long periods of inactivity. The bursty dynamics is characterized by the heavy tailed distribution of inter-event times $\tau$, defined as the time interval between consecutive events, shows a power-law decay as $P(\tau)\sim\tau^{-\alpha}$ with $\alpha\approx 0.7$ or $1$ for the MPC~\cite{Karsai2011} or for the email~\cite{Barabasi2005}, respectively. Two mechanisms for the origin of burstiness have been suggested: a) inhomogeneity due to the human circadian and weekly activity patterns~\cite{Malmgren2008,Malmgren2009a} and b) inhomogeneity rooted in the human task execution~\cite{Barabasi2005,Vazquez2006}. Although such dynamic inhomogeneity is obviously affected by the circadian and weekly patterns, it was claimed that the burstiness turns out to be robust with respect to the removal of circadian and weekly patterns from the time series of MPC and SM activities~\cite{Jo2011}. Here we will concentrate on considering the dynamic inhomogeneities other than those due to circadian and weekly patterns, namely due to those related to individual behavior.

In relation to the inhomogeneity of human task execution, several priority-based queuing models have been studied~\cite{Barabasi2005,Vazquez2006,Oliveira2009,Min2009,Cho2010,Min2011}. Each individual is assumed to have a task list of finite size and select one of tasks under the selection protocol, such as selecting the task with the largest priority. Most of these models focus on the waiting time of task, which is defined as the time interval between the arrival time and the execution time of the task. However, in some cases, since the arrival times of tasks to the queue are not given, the waiting times cannot be empirically measured and thus cannot be directly compared to the empirical inter-event times. Furthermore, in spite of studying the communication patterns, such as the email correspondence, the interaction between individuals has not been properly considered in the models~\cite{Barabasi2005,Vazquez2006}. Some interactive models defined on networks assume that the underlying networks are binary and fixed~\cite{Oliveira2009,Min2009,Cho2010,Min2011}. However, in reality both the topology and the weights of social networks co-evolve according to the individual task executions as well as to social interaction by cyclic and focal closure mechanisms.

Both the structural inhomogeneity of social interaction and the dynamical inhomogeneity of human individual behavior affect the dynamical processes taking place on evolving social networks. For example it has been shown that the Granovetter-type weight-topology coupling slows down information spreading~\cite{Onnela2007a}. By using the analogy between link weight and information-bandwidth information turns out to spread fast and to get trapped within communities due to the internal strong links (broad bandwidth) and the weak links (narrow bandwidth) between communities, respectively. In addition to the effect of Granovetter-type community structure on information spreading individual bursty behavior also plays a crucial role in social dynamics. The long inactive periods represented by large inter-event times, inhibit the information spreading compared to the randomized null model, while the bursty periods of short inter-event times do not necessarily enhance the spreading~\cite{Karsai2011}. Thus both the weight-topology coupling and the individual bursty dynamics should be taken into account and implement to a model in order to better understand the dynamics in the evolving social networks. 

The observation of Granovetter-type community structure and individual bursty dynamics calls for integrating both structural and dynamical inhomogeneities into single framework or model in order to better understand the social dynamics with the smallest set of parameters. Although there are some approaches in integrating these structural and dynamical properties, the bursty nature of human behavior has been inherently assumed in these models~\cite{Stehle2010,Zhao2011}. Instead, we are interested in the emergence of burstiness from the intuitive and natural model rules while at the same time generating the Granovetter-type community structure. In order to investigate the basic mechanisms responsible for various empirical observations, we incorporate the task handling process to the weighted network formation studied by Kumpula~\textit{et al.}~\cite{Kumpula2007}. In our model the weight assigned to a link is interpreted as the aggregate number of events on that link. Driven by both the cyclic and focal closure mechanisms a link is created by the first event occurring between individuals. Once created, the link is maintained by a series of events on that link, and finally removed by accidental memory loss of the individual. Each individual may initiate events or respond to those initiated by others, depending on the protocols determining the selection and execution of tasks given to individuals. 

Our model can be called co-evolutionary, in the sense that the task handling process of individuals affects the network evolution while the network structure constrains the individual behaviors. One of the typical issues in the co-evolutionary networks is that the timescale of network evolution competes with that of the dynamical process on the network~\cite{Holme2006,Iniguez2009}. In social dynamics the timescale for social relationship updates (a few weeks or months) is much larger than communication dynamics taking place on daily or hourly basis. In our case, since the events are the building blocks of the structure and the dynamics simultaneously, the relevant timescales are not explicitly controlled but emerged from the simple and intuitive rules of our model. In this paper, we show that by using the models with a few control parameters one can obtain the Granovetter-type community structure and also observe the emergence of bursty dynamics characterized by the heavy tailed inter-event time distribution.

This paper is organized such that we first introduce our two kinds of co-evolutionary models, the Triad-Interaction-enhanced model and the Process-Equalized model. Then we present the results for these models and discuss them in comparison with the empirical analysis results followed by the conclusions on the findings in the paper.

\section{Methods}

\begin{figure}[t]
\includegraphics[width=.7\columnwidth]{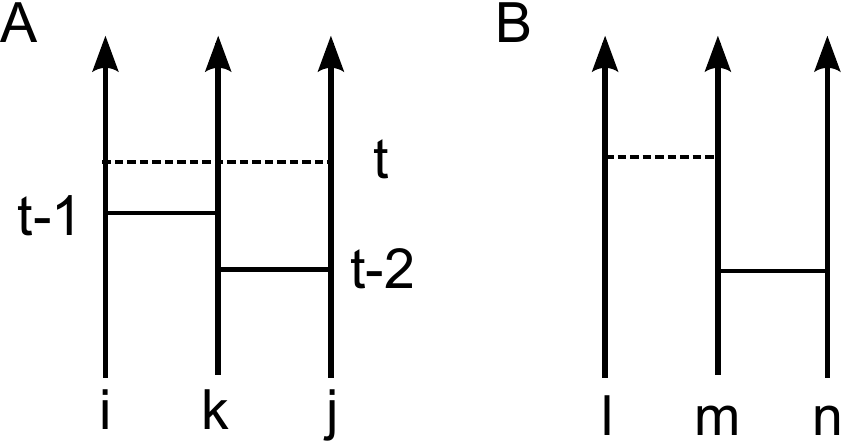}
\caption{{\bf Local and global attachments of the model.} Vertical lines represent the time lines of users. Horizontal solid and dashed lines represent events occurred and to be occurred between nodes, respectively. The number of events on a link defines the weight of the link. {\bf A}. Local Attachment (LA): The node $i$ has a chance to interact with node $j$ at time step $t$ only when there exists a temporal path connecting $i$ and $j$ through their common neighbor $k$ within time window $[t-2,t-1]$. {\bf B}. Global Attachment (GA): The isolated node $l$ has a chance to interact with a randomly chosen node $m$.}
\label{fig:model_LAGA}
\end{figure}

In our model we assume that the network evolves by means of link creation, link maintenance, and link deletion. Once a link between two stranger nodes is created by either the cyclic or the focal closure mechanisms, it is maintained by a series of events on that link, which we call the neighboring interaction (NI), or it is deleted by random memory loss. The focal closure mechanism is implemented by the random pairing of nodes, which is called global attachment or GA process. The cyclic closure mechanism is realized when a node interacts with its next nearest neighbor, which is called local attachment or LA process. While the GA process involves dyad interaction, the LA process is mediated by the third node, implying triad interaction. The NI process between neighboring nodes can happen directly, i.e. as dyad NI, or can be mediated by their common neighbor, i.e. as triad NI. Let us assume that only the event like the peer-to-peer phone call is considered. Then we can implement the triad interaction by splitting it into dyad interactions such that a node $i$ has a chance to interact with $j$ at time step $t$ only when both $i$ and $j$ have interacted with the third node $k$ recently, no more than, say, $2$ time steps before, see fig.~\ref{fig:model_LAGA}. In the following we propose two kinds of models. In the first kind the triad interaction takes place prior to the dyad interaction. We call this as Triad-Interaction-enhanced model (TI model in short). The TI model is a direct extension of the weighted network model by Kumpula~\textit{et al.}~\cite{Kumpula2007}, where the dyad NI process is analogous with Barab\'asi's task execution model~\cite{Barabasi2005}. In the second kind all the three processes (LA, GA, and NI) are considered equally and the framework of interacting and non-interacting tasks is adopted from~\cite{Oliveira2009}, as the variant of Barab\'asi's task execution model. We call this as Process-Equalized model (PE model in short).

Now let us consider an undirected weighted network with $N$ nodes. A weight of a link between nodes $i$ and $j$, denoted by $w_{ij}$, can be interpreted as the aggregate number of events between them. The number of neighbors of node $i$ is defined as the degree $k_i$. The time step of the most recent event between node $i$ and node $j$ is denoted by $t_{ij}$. Initially all nodes are set to be isolated, i.e. the initial network is without links.

\subsection{Triad-Interaction-enhanced model}

In the TI model the dynamics at each time step $t$ consists of the following three stages:

\textbf{1) Triad Interaction (LA and triad NI)}: For each pair of nodes $i$ and $j$ satisfying $\{t_{ik},t_{jk}\}=\{t-2,t-1\}$ with a third node $k$, we check whether $i$ and $j$ are connected. If they are connected, an event between $i$ and $j$ occurs, i.e. $w_{ij}\to w_{ij}+1$, corresponding to the triad NI process. Otherwise, the event between $i$ and $j$ occurred with probability $p_{_{LA}}$ leads to $w_{ij}=1$, implying a link creation by the LA process. These LA and triad NI processes are responsible for the community formation and weight reinforcement, respectively.

\textbf{2) Dyad Interaction (GA and dyad NI)}: Every node not involved in the previous stage selects a target node to make an event. If isolated, the node selects the target node from the whole population at random, preparing for the GA process. If non-isolated, the node selects the target node either from the whole population or from its neighbors with probabilities $p_{_{GA}}$ or $1-p_{_{GA}}$, respectively. In other words, all nodes are free to find new neighbors while the non-isolated nodes are also responsible for maintaining links to the existing neighbors, the degree of which is controlled by $p_{_{GA}}$. In the case of selecting the target from its neighbors, preparing for the dyad NI process, the probability of the node $i$ selecting its neighboring node $j$ is proportional to the weight between them, $w_{ij}$. Thus there is preference for the strong links. Targeting $j$ by $i$ is denoted by $i\to j$. The analogy between the target selection from the population or from the neighbors and the task selection from the task list will be discussed later.

The nodes having selected their targets make events with targets in a random order only when both the node and its target are not yet involved in any other event at this time step. If the node $i$ and its target $j$ were not connected, the event leads to a link creation between them, i.e. the realization of the GA process. Otherwise, the event between them results in $w_{ij}\to w_{ij}+1$, implying the dyad NI process. 

\textbf{3) Memory Loss}: With probability $p_{_{ML}}$, each node, $i$, becomes isolated and a stranger to all its neighbors $j$ as $w_{ij}=0$. This completes the time step $t$.

Through all the above stages it has been assumed that the target has no choice to reject the event initiated by some other node. We term this the OR protocol~\cite{Min2009} in a sense that it is enough for at least one of two nodes to initiate and make an event between them. Hence we call this version as the TI-OR model. Alternatively we can assume that an event can occur only in the reciprocal case, i.e. $i\to j$ and $j\to i$, which implies the AND protocol. It should be noted that for example a mobile phone user can reject a call from his/her friend by some reason. Here we will consider an TI-AND model, where the AND protocol is applied only to the dyad NI process.

\subsection{Process-Equalized model}

In the TI model, since the triad interaction is executed prior to the dyad interaction, one can not control the intensity of the triad interaction. Therefore we have devised the PE model where we consider the triad interaction on the equal footing with the dyad interactions, i.e. the LA, GA, and NI processes are equally considered. In this case we incorporate the task execution process with interacting and non-interacting tasks~\cite{Oliveira2009}, as described next.

Each node has the task list with one interacting task and one non-interacting task, denoted by $I$-task and $O$-task, respectively. The $I$-task represents the task requiring simultaneous interaction of two nodes, such as a phone call by a caller to a receiver, while the $O$-task represents some other task not requiring the simultaneity such as shopping, watching TV, etc. We count the inter-event times only for $I$-tasks, which settles down the issue of realistically interpreting the waiting time, as mentioned in~\cite{Oliveira2009}. The priorities of tasks are randomly drawn from the uniform distribution. 

In this model the dynamics takes place such that at each time step $t$, every node selected in a random order goes through the stages 1) and 2). Then the stage 3) is performed: 

\textbf{1) Task and Target Selection}: The node selects the task with larger priority. Only when it is $I$-task, this node, which we call a root node $i$, selects its target node either
\begin{itemize}
    \item from the whole population with probability $p_{_{GA}}$, i.e. the GA process, or
    \item from its next nearest neighbors with probability $p_{_{LA}}$, i.e. the LA process, or
    \item from its neighbors with probability $1-p_{_{GA}}-p_{_{LA}}$, i.e. the NI process.
\end{itemize}
For the LA process, the next nearest neighbor of the root node is defined as the node $j$ satisfying $\{t_{ik},t_{jk}\}=\{t-2,t-1\}$ with another node $k$. If the number of next nearest neighbors is more than $1$, one of them is selected at random. For the NI process, the probability to target one of the root node's neighbors $j$ is proportional to the weight $w_{ij}$, as in the TI model.

\textbf{2) Task Execution}: Only when the target node has not been involved in any event at this time, the event between the root node and the target node occurs, implying that the OR protocol is used. After this execution the priority of the $I$-task for the root node is replaced by the new random number while $j$'s task list is not updated, implying that the target node did not execute its $I$-task but simply responded to the root node.

\textbf{3) Memory Loss}: Each node becomes isolated with probability $p_{_{ML}}$, by which the time step $t$ is completed.

\subsection{Definitions of network properties}

We calculate various network properties for the numerically obtained networks. Given the weight distribution $P(w)$, the cumulative weight distribution is defined by 
\begin{equation}
    P_c(w)\equiv \int^\infty_w P(w')dw'.
\end{equation}
For each non-isolated node $i$, the number of next nearest neighbors, the individual clustering coefficient, and the strength are defined by
\begin{eqnarray}
    k_{nn,i} &\equiv& \frac{1}{k_i}\sum_{j\in \Lambda_i}k_j,\\
    c_i &\equiv& \frac{2 e_i}{k_i(k_i-1)},\\
    s_i &\equiv& \sum_{j\in \Lambda_i} w_{ij},
\end{eqnarray}
respectively. Here $\Lambda_i$ denotes the set of neighbors and $e_i$ denotes the number of links among the node $i$'s neighbors. The averages of the above quantities over the nodes with the same degree $k$ define the average number of next nearest neighbors $k_{nn}(k)$, the local clustering coefficient $c(k)$, and the average strength $s(k)$, respectively. In addition, to test the Granovetter-type community structure the overlap is defined for each link connecting nodes $i$ and $j$ as follows:
\begin{equation}
    O_{ij}\equiv \frac{|\Lambda_i \cap \Lambda_j|}{|\Lambda_i \cup \Lambda_j|},
\end{equation}
i.e. the fraction of the common neighbors over all neighbors of $i$ and $j$. The average over the links with the same weight $w$ defines the average overlap $O(w)$. For the dynamics we measure the inter-event time distributions $P(\tau)$.

\section{Results and Discussion}

The empirical analysis of mobile phone call data from a single operator in one European country for the first four months in 2007~\cite{Onnela2007b,Karsai2011} shows that $c(k)\sim k^{-\delta_c}$ with $\delta_c\approx 1$, $s(k)\sim k^{\delta_s}$ with $\delta_s\approx 1$. It also shows an increasing behavior of $k_{nn}(k)$, implying the assortativity, and an increasing behavior of $O(w)$ with slight decrease for very large $w$ values, where the increasing part implies the Granovetter-type community structure. Moreover, it was found that $P(\tau)\sim \tau^{-\alpha}$ with $\alpha\approx 0.7$. In addition the average degree $\langle k \rangle$ turned out to be around $3.0$ and the average clustering coefficient $\langle c \rangle$ around $0.13$ when the new year's day of 2007 is excluded. It should be noted that the average degree of the mobile phone call network extracted from the single operator dataset might be underestimated compared to the full mobile phone call network composed by many operators. Therefore, we assume that the overall  average degree of the whole social network is larger than $3$, i.e. around $10$. In this paper we consider the results to be relevant and comparable with reality, when $\langle k\rangle\approx 10$.

For the numerical simulations of the models described above we set the initial values as $N=5\times 10^4$ and $p_{_{ML}}=10^{-3}$ for all the cases considered. The simulations of these models are found to become stationary at about $t=3\times 10^3$, after which the numerical results are collected for $5\times 10^4$ time steps. 

\begin{figure}[t]
\includegraphics[width=\columnwidth]{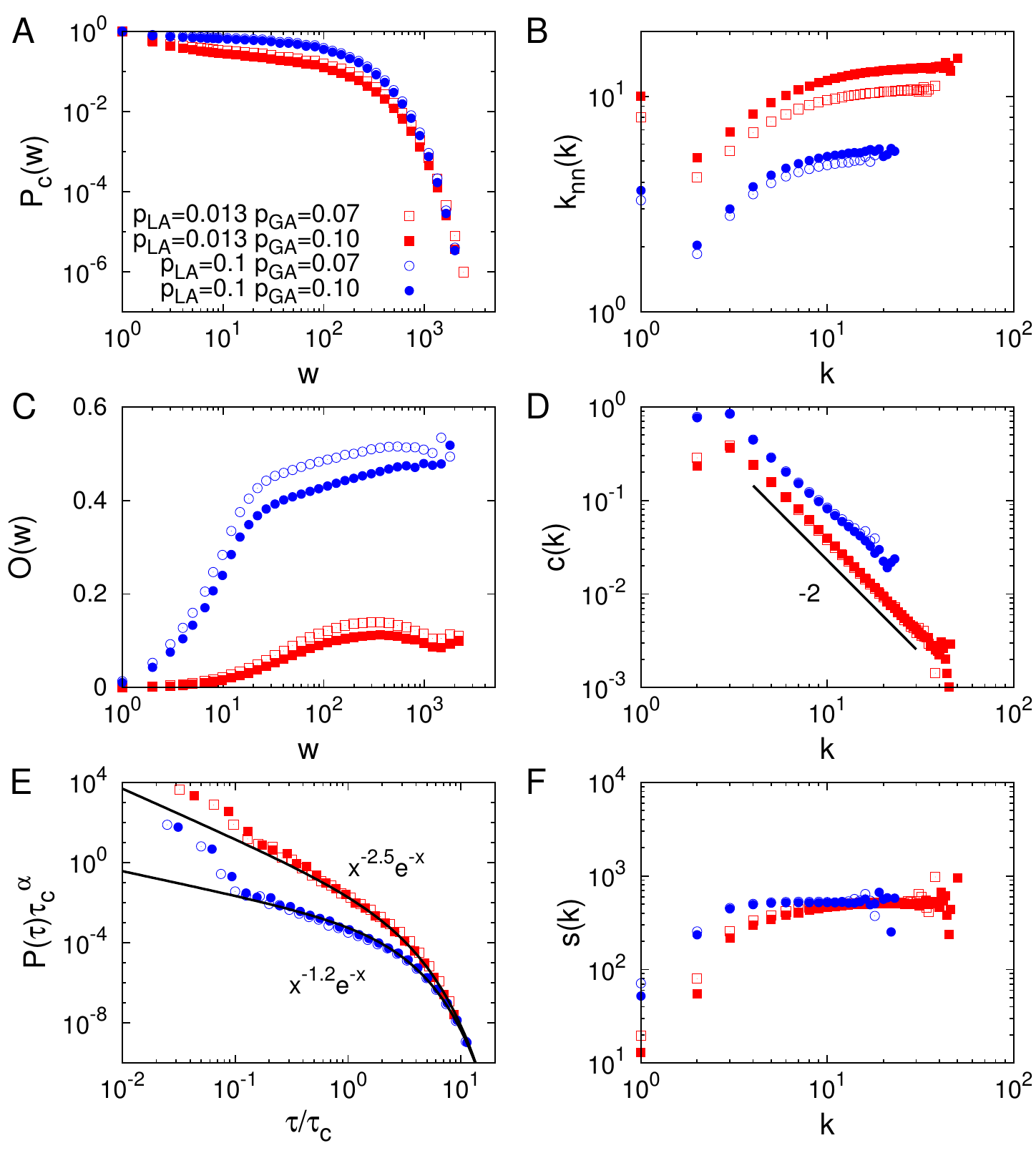}
\caption{{\bf TI-OR model.} {\bf A}. The cumulative weight distribution $P_c(w)$. {\bf B}. The average number of next nearest neighbors $k_{nn}(k)$. {\bf C}. The average overlap $O(w)$. {\bf D}. The local clustering coefficient $c(k)$. {\bf E}. The inter-event time distribution $P(\tau)$. {\bf F}. The average strength $s(k)$. Results are averaged over $50$ realizations for networks with $N=5\times 10^4$ and $p_{_{ML}}=10^{-3}$. We obtain $\langle k\rangle\approx 10.1$ and $\langle c\rangle\approx 0.08$ for $p_{_{LA}}=0.013$ and $p_{_{GA}}=0.1$. The cases with $p_{_{LA}}=0.1$ and/or with $p_{_{GA}}=0.07$ are also plotted for comparison.}
\label{fig:TIOR}
\end{figure}

\begin{figure}[t]
\includegraphics[width=\columnwidth]{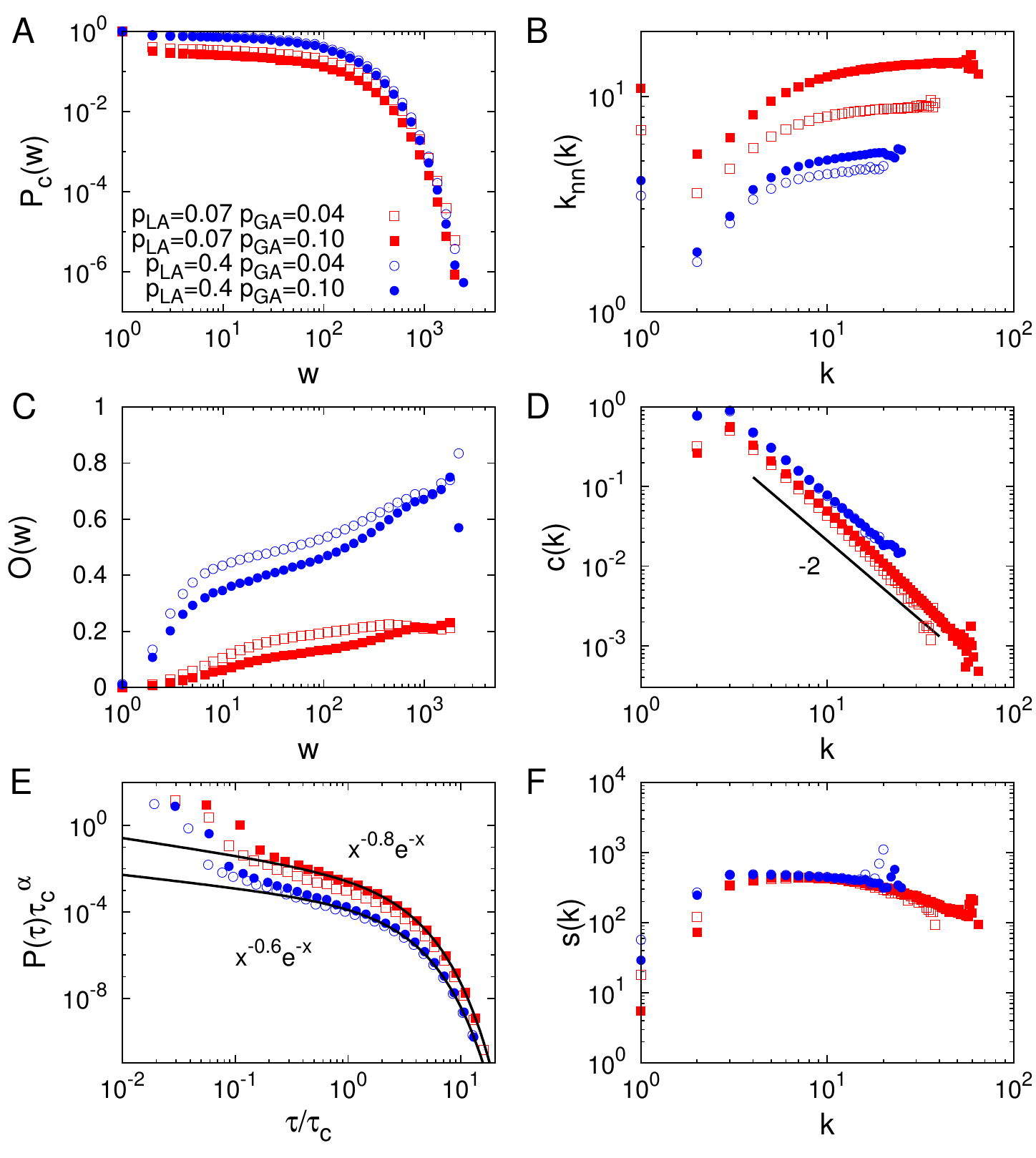}
\caption{{\bf TI-AND model.} {\bf A}. The cumulative weight distribution $P_c(w)$. {\bf B}. The average number of next nearest neighbors $k_{nn}(k)$. {\bf C}. The average overlap $O(w)$. {\bf D}. The local clustering coefficient $c(k)$. {\bf E}. The inter-event time distribution $P(\tau)$. {\bf F}. The average strength $s(k)$. Results are averaged over $50$ realizations for networks with $N=5\times 10^4$ and $p_{_{ML}}=10^{-3}$. We obtain $\langle k\rangle\approx 9.6$ and $\langle c\rangle\approx 0.13$ for $p_{_{LA}}=0.07$ and $p_{_{GA}}=0.1$. The cases with $p_{_{LA}}=0.4$ and/or with $p_{_{GA}}=0.04$ are also plotted for comparison.}
\label{fig:TIAND}
\end{figure}

\begin{figure}[t]
\includegraphics[width=\columnwidth]{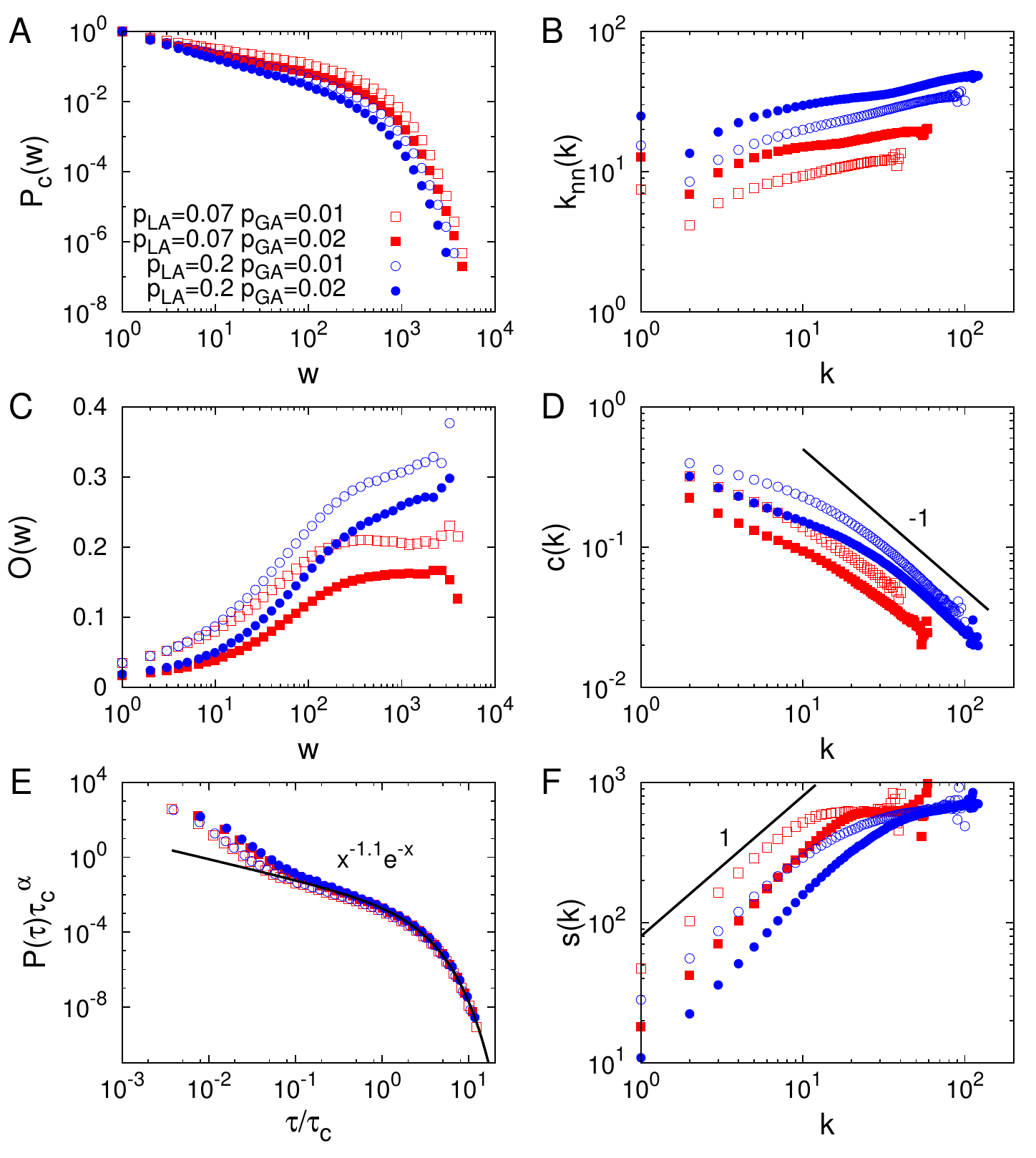}
\caption{{\bf PE model.} {\bf A}. The cumulative weight distribution $P_c(w)$. {\bf B}. The average number of next nearest neighbors $k_{nn}(k)$. {\bf C}. The average overlap $O(w)$. {\bf D}. The local clustering coefficient $c(k)$. {\bf E}. The inter-event time distribution $P(\tau)$. {\bf F}. The average strength $s(k)$. Results are averaged over $50$ realizations for networks with $N=5\times 10^4$ and $p_{_{ML}}=10^{-3}$. We obtain $\langle k\rangle\approx 9.9$ and $\langle c\rangle\approx 0.11$ for $p_{_{LA}}=0.07$ and $p_{_{GA}}=0.02$. The cases with $p_{_{LA}}=0.2$ and/or with $p_{_{GA}}=0.01$ are also plotted for comparison.}
\label{fig:PE}
\end{figure}

\begin{figure}[t]
\includegraphics[width=\columnwidth]{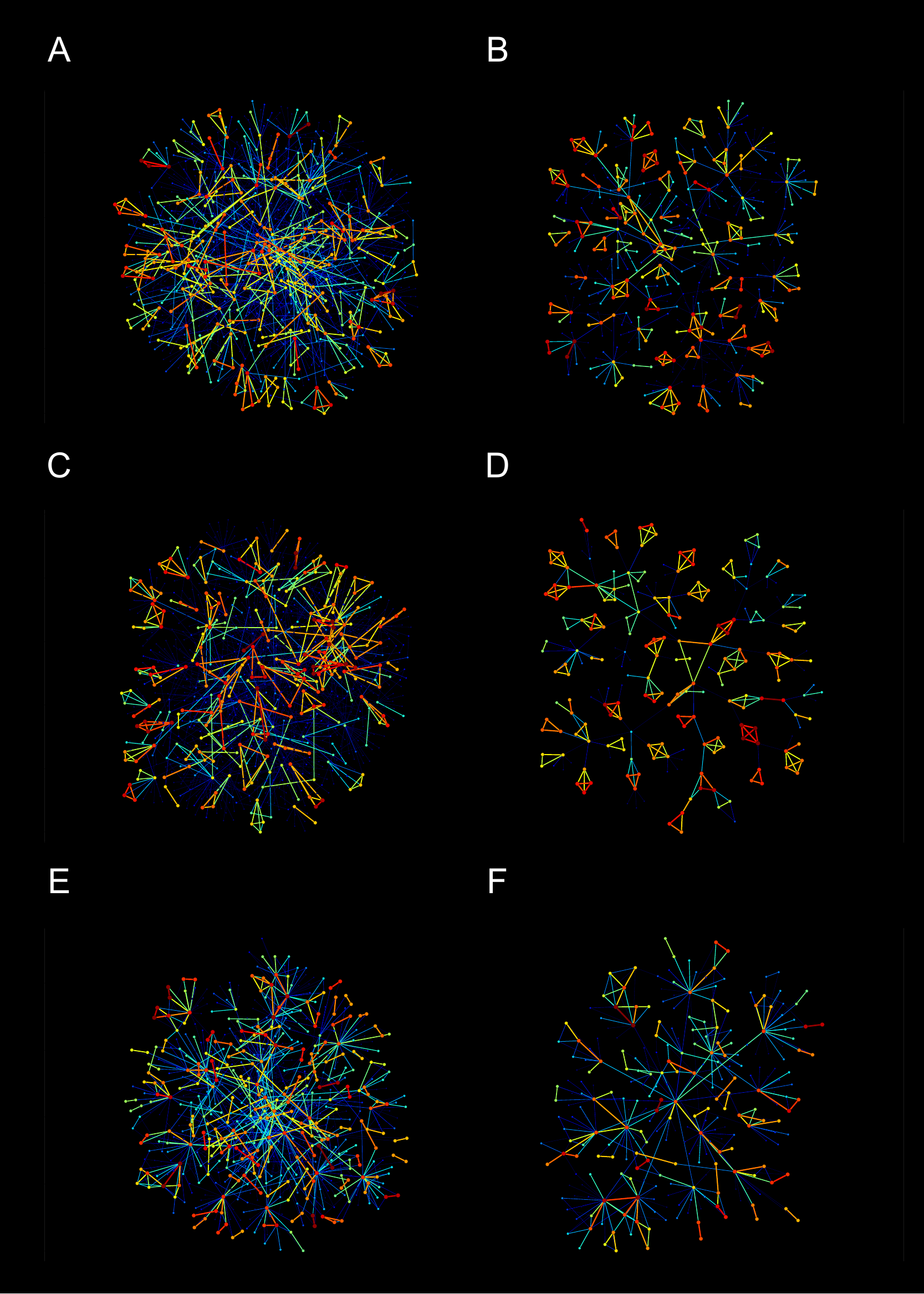}
\caption{{\bf Snowball samples~\cite{Lee2006} of networks for TI-OR model (A, B), for TI-AND model (C, D) and for PE model (E, F).} For each model, we plot the original sample starting from a random node ({\bf A}, {\bf C}, {\bf E}) and the one without the links with $w=1$ ({\bf B}, {\bf D}, {\bf F}) for the clear visualization. The color of links ranges from blue for weak links through yellow for intermediate links to red for strong links.}
\label{fig:snowball}
\end{figure}

\begin{figure}[t]
\includegraphics[width=\columnwidth]{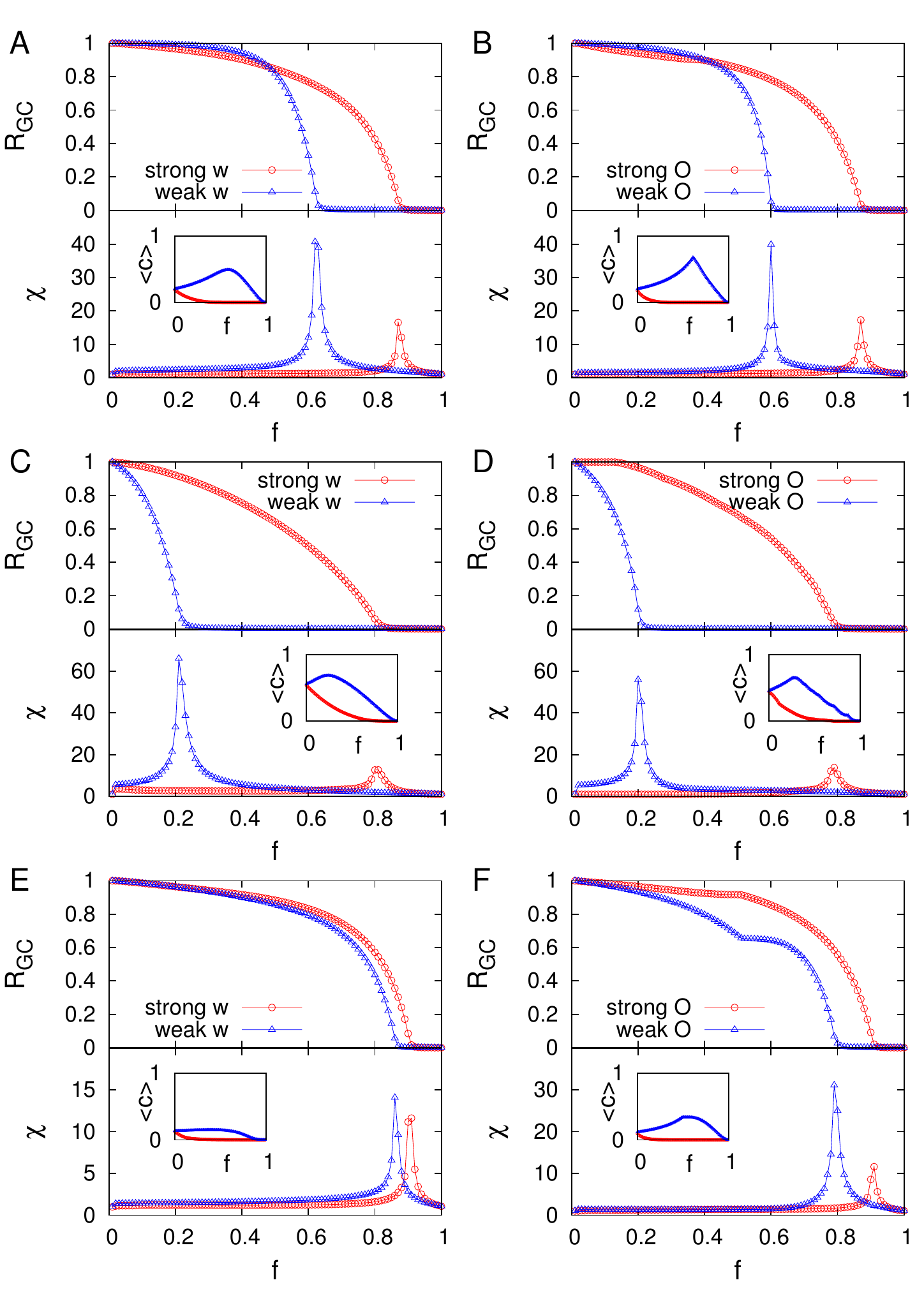}
\caption{{\bf Link percolation analysis for TI-OR model (A, B), for TI-AND model (C, D), and for PE model (E, F).} {\bf A, C, E}. The weight is used as the link strength. {\bf B, D, F}. The overlap is used as the link strength. For each panel, we calculate the fraction of giant component $R_{GC}$, susceptibility $\chi$, and clustering coefficient $\langle c\rangle$ (inset) as a function of the fraction of removed links, $f$. Results are averaged over $50$ realizations for networks originally with $\langle k\rangle\approx 10$ for each model.}
\label{fig:percolate}
\end{figure}

\subsection{Triad-Interaction-enhanced model}

For both the TI-OR and TI-AND models, we find that the cumulative weight distributions are broad but do not follow power-law behavior for the various values of $p_{_{LA}}$ and $p_{_{GA}}$ as in the empirical analysis, see figs.~\ref{fig:TIOR} and~\ref{fig:TIAND}. The similar behavior is found for degree and strength distributions (not shown). It turns out that as the empirical results, the networks are assortative and have the Granovetter-type community structure, characterized by the increasing behavior of $k_{nn}(k)$ for $k\geq 2$ and $O(w)$, respectively. Here most nodes with $k=1$ are supposed to be connected to randomly chosen nodes by the GA process, implying that $k_{nn}(1)\approx \langle k\rangle$. The sample networks shown in fig.~\ref{fig:snowball}{\bf A}-{\bf D} also confirm the emergence of Granovetter-type community structure, such that the communities of internal strong links are connected by weak links. In addition, for the TI-OR model with $p_{_{LA}}=0.1$ we observe a slightly decreasing behavior of $O(w)$ for large $w$ values, implying the existence of smaller but stronger communities. The decreasing behavior of the overlap was observed in the empirical analysis but not in the previous model studies~\cite{Kumpula2007}.

Based on the above observations it seems that a node is a member of a few strong triangles and connected to some other nodes outside its own triangles. This explains our finding of $\delta_c\approx 2$, different from the empirical result. It is because if the degree of a node increases mainly by means of the GA process, the number of links between neighbors remains while the number of all possible links grows as $k^2$, resulting in $c(k)\sim k^{-2}$. We also find $\delta_s\approx 0$ differently from the empirical value, which we will discuss later in relation to the dynamics.

In order to confirm the Granovetter-type community structure of networks, we perform the link percolation analysis. If links within communities are strong whereas links between them are weak as found in the empirical studies~\cite{Onnela2007a,Onnela2007b}, the network should disintegrate faster when the weak links are removed first than when the strong links are removed first. Note that as shown in fig.~\ref{fig:snowball} the links with weight $1$ form an apparently random network as backgrounds for the community structure. Thus we apply the link percolation to the giant components of networks without links with $w = 1$ and denote its size by $N'$. By removing links in an ascending or descending order of weights, we measure the remaining fraction of the giant component $R_{GC}$, the susceptibility $\chi$, and the average clustering $\langle c\rangle$ as a function of the fraction of removed links, $f$. Here the susceptibility is defined as $\chi=\sum n_s s^2/N'$, where $n_s$ denotes the number of clusters with size $s$ and the giant component is excluded from the summation. For the weak-link-first-removal cases we find the sudden disintegration of networks at the finite value of $f$, i.e. $f_c=0.62$ ($0.21$) for TI-OR (TI-AND) model. When the strong links are removed first, there is an apparent transition at the larger value of $f_c=0.87$ ($0.81$) for TI-OR (TI-AND) model as shown in fig.~\ref{fig:percolate}{\bf A} and {\bf C}. For the weak-link-first-removal cases the values of $f$ maximizing $\langle c\rangle$, denoted by $f_{max}$, are quite close to those of $f_c$. When using the overlap instead of the weight for the link percolation, almost the same behavior is observed in fig.~\ref{fig:percolate}{\bf B} and {\bf D} because $O(w)$ turns out to be the monotonically increasing function of $w$ in our model. 

For the temporal dynamics the inter-event time distributions are characterized by the power-law with an exponential cutoff, i.e. $P(\tau)\sim \tau^{-\alpha}\exp(-\tau/\tau_c)$, where the scaling regimes span over about one decade, see figs.~\ref{fig:TIOR}{\bf E} and~\ref{fig:TIAND}{\bf E}. In case of TI-OR model, $\alpha\approx 2.5$ or $1.2$ when $p_{_{LA}}=0.013$ or $0.1$, respectively. In the case of TI-AND model, when $p_{_{LA}}=0.07$ or $0.4$, we find $\alpha\approx 0.8$ or $0.6$, respectively, both of which are close to the empirical value $0.7$ of MPC dataset within error bars. In all cases, the values of $\alpha$ are smaller for larger values of $p_{_{LA}}$ but are barely affected by the value of $p_{_{GA}}$. The values of $\tau_c$ turn out to be larger for larger values of $p_{_{LA}}$ and for smaller values of $p_{_{GA}}$. The maximum value of $\tau_c$ is around $50$.

To figure out what are the possible underlying mechanism for these findings, we first identify the triangular chain interaction (TCI) among three neighboring nodes, say $i$, $j$, and $k$: Both the event between nodes $i$ and $j$ at time step $t-2$ and the event between nodes $j$ and $k$ at time $t-1$ lead to an event between nodes $i$ and $k$ at time $t$, again leading to another event between nodes $i$ and $j$ at time $t+1$ and so on, unless interrupted either by the events from/to nodes outside the triangle or by a random memory loss of nodes in the triangle. Since the TCI is exclusive due to the priority of the triad interaction including the LA process, the LA process enhanced by the large value of $p_{_{LA}}$ inhibits the interruption by the events from/to nodes outside the triangle, including the GA process, and thus making the community structure more compact in turn resulting in a smaller average degree. In case of TI-OR model with $p_{_{GA}}=0.1$, $\langle k\rangle\approx 10.1$ or $4.2$ for $p_{_{LA}}=0.013$ or $0.1$, respectively. While the compact community structure enhances the TCI again, explaining the observed peaks of $P(\tau)$ at $\tau=1$ and $2$, it can also make some neighbors of the TCI nodes wait for long time to interact with the TCI nodes. Hence, the larger value of $p_{_{LA}}$ gives rise to larger fluctuation in the inter-event times, implying a smaller value of the power-law exponent $\alpha$ and a larger value of the cutoff $\tau_c$, as observed. Based on this argument, the effect of $p_{_{LA}}$ dominates over that of $p_{_{GA}}$, so that the value of $p_{_{GA}}$ barely affects the scaling of inter-event time distributions but it controls the value of $\tau_c$. The larger value of $p_{_{GA}}$ allows nodes to choose a random target and thus interrupt the inter-event times of targets more frequently, leading to a smaller value of $\tau_c$. The numerical results in the case of the TI-AND model can be explained by the same arguments, except for the observed values of $\alpha$ less than those found in the case of the TI-OR model. Note that in general the AND protocol inhibits the possibility of events.

The heavy tailed distribution of inter-event times, i.e. bursty dynamics, was not expected but it emerged from the model. Analogously with the task execution model suggested by Barab\'asi~\cite{Barabasi2005}, the dyad NI process can be interpreted such that a node $i$ has the task list with size $k_i$ and it selects one of neighbors (tasks) $j$ with probability proportional to the priority of the task, i.e. the weight $w_{ij}$ in our model. The degree $k_i$ also varies depending on the link creation and deletion processes. A node having been isolated by the memory loss tries to interact with strangers. Once being connected to some other node by the GA process, its degree increases partly by means of the LA process but it will not diverge. The degree mostly fluctuates and sometimes remains unchanged for long periods of time. And the node finally becomes isolated again by the memory loss. Thus, the whole life-cycle of a node is assumed to consist of two types of periods, i.e. one with fixed-size and the other with variable-size task list. The periods of fixed-size task list, i.e. fixed degrees, are up to several hundred time steps, which are much larger than the observed $\tau_c$. This implies the natural separation of timescales between network change and dynamics on the network, which is consistent with everyday experience of mobile phone usage. Due to the timescale separation the inter-event time distribution for the whole period can be represented by the superposition of those for fixed-size period and for variable-size period. Thus, to understand the effect of size variability on the scaling behavior of bursty dynamics, we refer to the previous works studied in the different kinds of models, such as by V\'azquez~\textit{et al.}~\cite{Vazquez2006}. When the task list has a variable (fixed) size in the Barab\'asi model, the power-law exponent for the waiting time distribution turns out to be $3/2$ ($2$). According to the argument that the distribution of the inter-event times derived from the waiting times has the same power-law exponent as that of the waiting times, one can expect the similar values of exponent from our model. However, this is not the case with our model, so we leave this for the more rigorous analysis in the future. 

Finally, the apparent overall independence of the average strength $s(k)$ on $k$ for large values of $k$ is attributed to the fact that once the node is a member of the TCI, its activity becomes effectively independent of its degree due to the exclusive property of TCI. We observe even the decreasing behaviors of $s(k)$ for the larger $k$ values in the TI-AND model, i.e. the AND protocol based interaction with too many neighbors can make nodes failing to interact with any neighbors.

\subsection{Process-Equalized model}

The TI models show the expected behaviors of Granovetter-type community structure and the heavy tailed inter-event time distribution but they do not yield the expected behavior of the local clustering coefficient and average strength of the nodes. This is mainly due to too strong effect of the triad interaction and that is why we need to consider the PE model for modeling improvement and comparison with empirical results.

With the PE model we find that the cumulative weight distributions $P_c(w)$ are broad, that the overlap $O(w)$ increases with $w$, i.e. showing Granovetter-type community structure, that the average number of next nearest neighbor $k_{nn}(k)$ increases with $k$, i.e. showing the network being assortativity for $k\geq 2$, and that $c(k)\sim k^{-\delta_c}$ with $\delta_c\approx 1$ and $s(k)\sim k^{\delta_s}$ with $\delta_s\approx 1$, as shown in fig.~\ref{fig:PE}. All these results are consistent with the empirical analysis on real data. Based on the sample networks in fig.~\ref{fig:snowball}{\bf E} and {\bf F}, it is evident that the TCI becomes weaker and less exclusive than in the case of the above TI models. Therefore, as the degree of a node increases, the neighbors of that node have the increasing chance to interact with each other, resulting in $c(k)\sim k^{-1}$.

In fig.~\ref{fig:percolate}{\bf E} and {\bf F} we show the results of the link percolation analysis, done to confirm Granovetter-type community structure. We find that when the weak links are removed first, the percolation transition occurs at $f_c=0.86$. On the other hand when the strong links are removed first, a transition is observed at $f_c=0.91$, implying that the strong links play the role of bridges between communities. This is also evident in the sample networks in fig.~\ref{fig:snowball}{\bf E} and~{\bf F}. The curve of the average clustering coefficient $\langle c\rangle$ turns out to be flat for a wide range of $f$ values. Similar behaviors are also observed when the overlap is used instead of the weight in the link percolation analysis. For the weak-link-first-removal we find $f_c=0.79$ and $f_{max}=0.55$, where yet another kink in the curve of $R_{GC}$ is observed. This implies that the network goes through two abrupt changes, first at $f_{max}$ and then at $f_c$.

Here we also observe the heavy tailed distributions of inter-event times with exponential cutoffs following a power law behavior with the exponent of $\alpha\approx 1.1$. The task execution model for each node would result in $\alpha=1$ as in the case of Barab\'asi's queuing model if only initiating the $I$-tasks are counted as the relevant events and if the neighbors of the node always respond to that node. However, the nodes are supposed to interact with each other such that by initiating $I$-tasks some root nodes can interrupt the inactive periods of their target nodes, which in general decreases the inter-event times. On the other hand, if the target is already involved in another event so that the trials by the root nodes fail, the inter-event times of corresponding root nodes would increase up to the points of next successful events occurring. The observed value of $\alpha\approx 1.1$ indicates that any of the mentioned factors did not affect much the scaling behavior of the distributions. The values of $\tau_c$ are largely or barely affected by the value of $p_{_{GA}}$ or $p_{_{LA}}$, respectively, in an anti-correlated way. The maximum value of $\tau_c$ is around $270$.

The observation of the average strength $s(k)\sim k$ behavior can be explained by considering the dynamics where the OR protocol is adopted. In this case the nodes with many neighbors might receive more calls from their neighbors than those with few neighbors do, while the chance to make calls is the same for any node.

\subsection{Conclusions}

We have studied the emergence of Granovetter-type community structure, characterized by the increasing behavior of overlap as a function of the link weight, and the heavy tailed inter-event time distributions, i.e. bursty dynamics, in a single framework of simultaneously evolving weighted network model. By incorporating simple and intuitive task execution models for human dynamics into the weighted network model reproducing the Granovetter-type community structure of social systems, we successfully observe the qualitatively same behaviors as observed in the empirical networks based on the mobile phone call (MPC) dataset. In addition, we have found that the exclusive triangular chain interaction (TCI) identified in the TI models plays the central role both in community structure formation and bursty dynamics. For the existence of TCI we have the evidence from the empirical study on the dynamic motifs of MPC communication~\cite{Kovanen2009}. The numerical results from TI-OR and TI-AND models are qualitatively the same except for the power-law exponent $\alpha$ of inter-event time distributions. The values of $\alpha$ from TI-AND model turn out to be closer to the empirical value $0.7$ for the MPC, implying that the AND protocol is necessary to properly model the MPC communication. Furthermore, in the PE model, by relaxing the exclusive property of TCI to some extent we could obtain more realistic results at least for the network structure. The scaling behavior of inter-event time distributions seems to be mainly affected by the incorporated framework of interacting and non-interacting tasks, which should be made clear in the future.

Finally we believe that building simple empirical-observation-based models, like our TI- and PE-models, by incorporating the process of human task execution by priority-based queuing with the basic processes of friendship-network formation by cyclic and focal closure mechanisms enable us to better understand the underlying mechanisms of real co-evolutionary networks. Furthermore, these models enable us to explore the social dynamics in these networks as done differently by Karsai~\textit{et al.}~\cite{Karsai2011} with the susceptible-infected (SI) dynamics for the mobile phone call communication. Moreover, the scaling properties and finite-size scaling of real networks are usually not so informative but can be considered and made more informative by means of simple but still quite realistic models, where one can control the system size and other parameters as well.     

\begin{acknowledgments}
Financial support by Aalto University postdoctoral program (HJ), by the Academy of Finland, the Finnish Center of Excellence program 2006-2011, project no. 129670 (RKP, KK) are gratefully acknowledged.
\end{acknowledgments}

\bibliography{burstComm}
\end{document}